\documentclass[fleqn,usenatbib]{mnras}

\usepackage{newtxtext,newtxmath}
\usepackage{orcidlink}

\usepackage[T1]{fontenc}

\DeclareRobustCommand{\VAN}[3]{#2}
\let\VANthebibliography\thebibliography
\def\thebibliography{\DeclareRobustCommand{\VAN}[3]{##3}\VANthebibliography}

\usepackage{graphicx}	
\usepackage{amsmath}	

\title[Super-Eddington Accretion to Form GW190425]{Super-Eddington Accretion as a Possible Scenario to Form GW190425}

\author[Zhang et al.]{W. T.\, Zhang,$^{1,9}$
Z. H. T.\, Wang,\orcidlink{0000-0001-9592-6671}$^{2,9}$
J.-P.\, Zhu,\orcidlink{0000-0002-9195-4904}$^{3,4}$
R.-C.\, Hu,\orcidlink{0000-0002-6442-7850}$^{5}$
X. W.\, Shu,$^{1}$
Q. W.\, Tang,\orcidlink{0000-0001-7471-8451}$^{6}$
S. X.\, Yi,\orcidlink{0000-0003-0672-5646}$^{7}$
\newauthor
F.\, Lyu,\orcidlink{00000-0002-6072-3329}$^{8}$
E. W.\, Liang,\orcidlink{0000-0002-7044-733X}$^{2}$
and Y.\, Qin\orcidlink{0000-0002-2956-8367}$^{1,2}$\thanks{E-mail: yingqin2013@hotmail.com}\\
$^{1}$Department of Physics, Anhui Normal University, Wuhu, Anhui 241002, China\\
$^{2}$Guangxi Key Laboratory for Relativistic Astrophysics, School of Physical Science and Technology, Guangxi University, Nanning 530004, China\\
$^{3}$School of Physics and Astronomy, Monash University, Clayton Victoria 3800, Australia\\
$^{4}$OzGrav: The ARC Centre of Excellence for Gravitational Wave Discovery, Australia\\
$^{5}$Department of Physics and Astronomy, University of Nevada, Las Vegas, NV 89154, USA\\
$^{6}$Department of Physics, School of Physics and Materials Science, Nanchang University, Nanchang 330031, China\\
$^{7}$School of Physics and Physical Engineering, Qufu Normal University, Qufu, Shandong 273165, China\\
$^{8}$Astronomical Research Center, Shanghai Science \& Technology Museum, Shanghai 201306, China \\
$^{9}$ These authors contributed equally to this work.
}
\date{Accepted 2023 September 13. Received 2023 September 10; in original form 2023 July 31}

\pubyear{2023}

\begin{document}
\label{firstpage}
\pagerange{\pageref{firstpage}--\pageref{lastpage}}
\maketitle

\begin{abstract}
On 2019 April 25, the LIGO/Virgo Scientific Collaboration detected a compact binary coalescence, GW190425. Under the assumption of the binary neutron star (BNS), the total mass of $3.4^{+0.3}_{-0.1}\, M_\odot$ lies five standard deviations away from the known Galactic population mean. In the standard common envelope scenario, the immediate progenitor of GW190425 is a close binary system composed of an NS and a He-rich star. With the detailed binary evolutionary modeling, we find that in order to reproduce GW190425-like events, super-Eddington accretion (e.g., $1,000\,\dot{M}_{\rm Edd}$) from a He-rich star onto the first-born NS with a typical mass of 1.33 $M_\odot$ via stable Case BB mass transfer (MT) is necessarily required. Furthermore, the immediate progenitors should potentially have an initial mass of $M_{\rm ZamsHe}$ in a range of $3.0-3.5$ $M_\odot$ and an initial orbital period of $P_{\rm init}$ from 0.08 days to 0.12 days, respectively. The corresponding mass accreted onto NSs via stable Case BB MT phase varies from $0.70\, M_\odot$ to $0.77\, M_\odot$. After the formation of the second-born NS, the BNSs are expected to be merged due to gravitational wave emission from $\sim$ 11 Myr to $\sim$ 190 Myr.
\end{abstract}

\begin{keywords}
Gravitational waves -- stars: neutron -- binaries: close
\end{keywords}



\section{Introduction}
After the announcement of the first binary neutron star (BNS) merger \cite[GW170817;][]{gw170817}, the LIGO/Virgo Scientific Collaboration (LVC), on 2019 April 25, detected the second signal of a BNS merger GW190425 with a signal-to-noise ratio of 12.9 \citep{gw190425}. The binary GW190425, with a total mass of $3.4^{+0.3}_{-0.1}\, M_\odot$, lies five standard deviations away from the mean mass of known Galactic BNS systems \cite[][and references therein]{Farrow2019}. So far, the origin of GW190425 remains unclear, although different groups have independently diagnosed its formation path.

The canonical formation channels of BNS systems can be split into two distinct categories: (i) the isolated binary evolution channel \cite[e.g.,][]{Flannery1975,Zwart1998,Kalogera2007,Postnov2014,Shao2018,Vigna2018,Kruckow2018,Mapelli2018,Giacobbo2018,Giacobbo2019,Giacobbo2020}, and dynamical assembly \cite[see][and references therein]{Phinney1991,Grindlay2006,Benacquista2013,Palmese2017,Andrews2019,Ye2020}. The former is the dominant formation channel for Galactic BNS, in which the two NSs are formed in an isolated binary \citep{Tauris2017,Vigna2018}. Assuming the standard channel, the two NSs of GW190425 are formed in a sequence of supernova explosions. The immediate progenitor of GW190425 is a close binary system composed of an NS and a He-rich star, which can be the outcome of the classical isolated binary evolution through the common envelope. In the subsequent evolution, it might have involved a phase of stable or unstable Case BB MT (MT initiates during the shell-helium burning phase). Recently, \cite{Safarzadeh2020} employed population synthesis modeling to find that an unstable MT channel faces a challenge to reconcile the merger rate of GW190425. Subsequently, \cite{Zhu2020} constrained its eccentricity of $e$ $\leqslant$ 0.007 at 10 Hz with 90\% confidence, indicating no evidence for or against the unstable MT scenario. In the standard scenario, it was found in \cite{Vigna2018} that Case BB MT from He-rich stars onto NSs is most likely dynamically stable, broadly in agreement with the investigations in \cite{Tauris2015} and earlier studies in \cite{Ivanova2003}. Generally, massive He-rich stars tend to be compact \citep[see their Figure 1 in][]{Qin2023} and thus could avoid MT onto an NS. By performing three-dimensional simulations of the supernova explosion of massive ($\gtrsim 9\, M_\odot$) He-rich stars, \cite{Vigna2021} found that fallback could lead to significant mass growth in the second-born NS, providing a promising channel to explain the formation of massive BNS systems such as GW190425.

Theoretically, the MT rates for highly super-Eddington accretion are physically plausible \citep{King2001} for different compact objects. In the case of stellar-mass black hole (BH) X-ray binaries, \cite{Moreno2011} pointed out that for M33 X-7 and LMC X-1, their inferred BH high spins had to be obtained through hypercritical accretion. Recently, \cite{Qin2022RAA} employed detailed binary evolution to find that the hypercritical accretion channel could well match the observed properties of Cygnus X-1. Additionally, super-Eddington accretion has also been recently studied to investigate the origin of BH high spins in binary BHs \cite[e.g.,][]{Bavera2021,Zevin2022,Shao2022,Briel2023}. With the population synthesis modeling, \cite{Wiktorowicz2015} showed that an NS accreting from a low-mass He-rich star in the thermal-timescale phase could have an MT rate of $\sim$ 10$^{-2}$ $M_{\odot}$  yr$^{-1}$. Until recently, observational evidence for super-Eddington accretion of NSs has been continuously accumulated \citep[see][a recent review]{Kaaret2017}. The most remarkable evidence for super-Eddington accretion comes from a recent detection of the ultraluminous X-ray (ULX) pulsar in NGC 5907 \citep{Israel2017}, which contains accreting NSs with an accretion rate apparently above the Eddington limit by factors up to $\sim$ 500 (isotropic peak luminosity of $\sim$ 1, 000 times the Eddington limit). Additionally, \cite{Ghodla2023} recently suggested that sustained super-Eddington accretion might always be possible around compact objects (i.e., NSs and BHs). More recently, \cite{Zhou2023} reported the first evidence for the presence of He-rich donor stars in ULXs, which is consistent with the prediction that He-rich donor stars might be common in ULXs. Motivated by these mentioned above, we here employ detailed binary evolution to investigate the potential formation path of GW190425, under the assumption of super-Eddington accretion from He-rich stars onto NSs.

In this paper, we first introduce the main methods in Section \ref{sec:2}. We then show in Section \ref{sec:3} the detailed evolution of low-mass single He-rich stars, including rotation and no rotation. In Section \ref{sec:4}, we present detailed binary modeling of He-rich stars in close orbits with NSs as companions under different Eddington accretion limits. Finally, the main conclusions and some discussion are summarized in Section \ref{sec:5}.

\section{Methods} \label{sec:2}
We used release 15140 of the \texttt{MESA} stellar evolution code \citep{Paxton2011,Paxton2013,Paxton2015,Paxton2018,Paxton2019,Jermyn2023} to perform the stellar structure and binary evolution calculations in this work. We adopted a metallicity of $Z = Z_{\odot}$, where the solar metallicity is $Z_{\odot} = 0.0142$ \citep{Asplund2009}. We first created pure single He-rich stars at the zero-age helium main sequence (ZAHeMS) following the same method \citep{Qin2018,Bavera2020,Hu2022,Hu2023,Fragos2023,Qin2023,lv2023} and further relaxed the created ZAHeMS to reach the thermal equilibrium, where the helium-burning luminosity just exceeds $99\%$ of the total luminosity. We modeled convection using the mixing-length theory \citep{MLT1958} with a mixing-length $\alpha_{\rm mlt}=1.93$. We adopted the Ledoux criterion to treat the boundaries of the convective zone and considered the step overshooting as an extension given by $\alpha_p = 0.1 H_p$, where $H_p$ is the pressure scale height at the Ledoux boundary limit. Semiconvection \citep{Langer1983} with an efficiency parameter $\alpha_{\sc}=1.0$ was adopted in our modeling. The network of \texttt{approx12.net} was chosen for nucleosynthesis. We treated rotational mixing and angular momentum transport as diffusive processes \citep{Heger2000}, including the effects of the Goldreich–Schubert–Fricke instability, Eddington–Sweet circulations, as well as secular and dynamical shear mixing. We adopted diffusive element mixing from these processes with an efficiency parameter of $f_c=1/30$ \citep{Chaboyer1992,Heger2000}. MT was modeled following the Kolb scheme \citep{Kolb1990} and the implicit MT method \citep{Paxton2015} was adopted. MT was assumed to be conservative for sub-Eddington accretion. However, for mass not accreted by the accretor, we assumed that the excess material is lost from the vicinity of the accretor as an isotropic wind carrying the specific angular momentum of the accretor.

We adopted the ``\texttt{Dutch}'' scheme \citep{Jager1988,Nugis2000,vink2001,Glebbeek2009} for the wind mass-loss of He-rich stars, calibrated by multiplying with a \texttt{Dutch$_{-}$scaling$_{-}$factor} = 0.667 as in \cite{Hu2022} to match the recently updated modeling of He-rich stars' winds \citep{Higgins2021}. We took into account the rotationally enhanced wind mass-loss \citep{Heger1998,Langer1998} as follows:
    \begin{equation}\label{ml}
    \centering
    \dot{M}(\omega)= \dot{M}(0)\left(\frac{1}{1-\omega/\omega_{\rm crit}}\right)^\xi,
    \end{equation}
where $\omega$ and $\omega_{\rm crit}$ ($\omega_{\rm crit}^2 = (1- L/L_{\rm Edd})GM/R^3$, $L_{\rm Edd}$ is the Eddington luminosity) are the angular velocity and critical angular velocity at the surface, respectively. The default value of the exponent $\xi = 0.43$ is taken from \citet{Langer1998}. No gravity darkening effect is accounted for \citep[see][for a discussion on the impact of this process]{Maeder2000}.

In the modeling of a binary system composed of an NS and a He-rich star, we evolved He-rich stars to reach the carbon depletion in the center and treated an NS as a point mass. We calculated the baryonic remnant mass following the ``\texttt{delayed}'' supernova prescription \citep{Fryer2012}, which can produce remnant mass in the mass gap ($\sim2.5 - 5\,M_\odot$) between NSs and BHs. In general, the ``\texttt{rapid}'' mechanism \citep{Fryer2012} predicts the mass gap that might be from X-ray binary observations, while both the ``\texttt{delayed}'' and stochastic prescription \citep{Mandel2020} can produce remnant mass in the mass gap. It was found that the stochastic prescription tends to produce similar distributions of NS mass with different companions when compared with the ``\texttt{delayed}'' mechanism \cite[see Figure 10 in][]{Shao2021}. Therefore, we expect that adopting the stochastic recipe will have no significant impact on our results. We took into account neutrino loss as in \cite{Zevin2020}. We assumed that the maximum NS mass is 2.2 $M_{\odot}$ \citep[taken from the constraints by the observations of Galactic pulsars and GW binaries, e.g.,][]{Margalit2017,Romani2022,Zhu2022Population,Abbott2023Population} in this work.

We followed the Kolb scheme \citep{Kolb1990} to model the MT and adopted the implicit MT method \citep{Paxton2015}. For He-rich stars, we use the dynamical tides, which are suitable for massive stars with radiative envelopes \citep{Zahn1977}. We calculated the timescale of synchronization following the prescription in \citep{Zahn1977,Hut1981,Hurley2002}. We also adopted the recently updated fitting formula of the tidal torque coefficient $E_2$ as in \cite{Qin2018}.

Given NS as an accreting object, the accretion luminosity is defined as
\begin{equation}\label{lum}
L = \eta  \dot{M}_{\rm acc} c^2,
\end{equation}
where $\dot{M}_{\rm acc}$ is the accretion rate and $c$ is the speed of light in vacuum. The radiative efficiency $\eta$ is given as
\begin{equation}\label{eta}
\eta = \frac{G M_{\rm acc}}{R_{\rm NS} c^2}, 
\end{equation}
where $M_{\rm acc}$ are $R_{\rm NS}$ are the accreting-object mass and radius ($\rm NS$ mass and radius in this work), respectively. For NS, we adopted a constant $R_{\rm acc}$ of 12.5 km \citep{Most2018,Miller2019,Riley2019,Landry2020,gw190425,Kim2021,Biswas2021,Raaijmakers2021}.
The Eddington luminosity (also referred to as the Eddington limit) is the maximum luminosity at which the force of radiation acting outward balances the gravitational force acting inward. We calculated the Eddington luminosity $L_{\rm Edd}$ for NS as an accretor following the standard formulae \citep{Frank2002} as follows:
\begin{equation}\label{ledd}
L_{\rm Edd} = \frac{4\pi G M_{\rm acc} c }{\kappa},
\end{equation}
where $G$ is the gravitational constant and $\kappa$ is the opacity contributing from pure electron scattering, i.e. $\kappa$ = 0.2(1 + X) $\rm cm^2$ $\rm g^{-1}$. For the accretion of helium onto a $\rm NS$, the hydrogen mass fraction $\rm X$ = 0. Combining the equations \ref{lum} and \ref{ledd} gives the Eddington mass-accretion rate:
\begin{equation}\label{m_acc}
\dot{M}_{\rm acc} =  \frac{4\pi G M_{\rm acc} c }{\kappa c \eta}.
\end{equation}

As the accretion onto the NS increases, the radiation pressure continues building up near its surface, which resists the inflow in excess of the Eddington accretion rate. Under this circumstance, a common envelope may develop provided that the trapping radius exceeds the Roche lobe radius of the $\rm NS$. The trapping radius \citep{Begelman1979,King1999,Ivanova2003} is the boundary at which the luminosity generated by infalling material on the accretor is equal to the Eddington accretion limit, i.e.,
\begin{equation}\label{eta}
R_{\rm trap} = \frac{\dot{M_{\rm tr}}}{\dot{M_{\rm Edd}}} \frac{R_{\rm NS}}{2}, 
\end{equation}
where $\dot{M}_{\rm tr}$ is the MT rate onto NS, $\dot{M}_{\rm Edd}$ ($\dot{M}_{\rm Edd} = L_{\rm edd} /c^2$) is the Eddington accretion rate, and $R_{\rm NS}$ the NS radius, respectively. As shown in \cite{Ivanova2003}, the above criterion can be reexpressed in terms of a critical MT rate ($\dot{M}_{\rm crit}$, see their equation 15). Here, we assume that a common envelope may form if the MT rate approaches $\dot{M}_{\rm crit}$. 

\section{Modeling of low-mass single He-rich stars with and without rotation} \label{sec:3}
\begin{figure}
     \includegraphics[width=1.0\columnwidth]{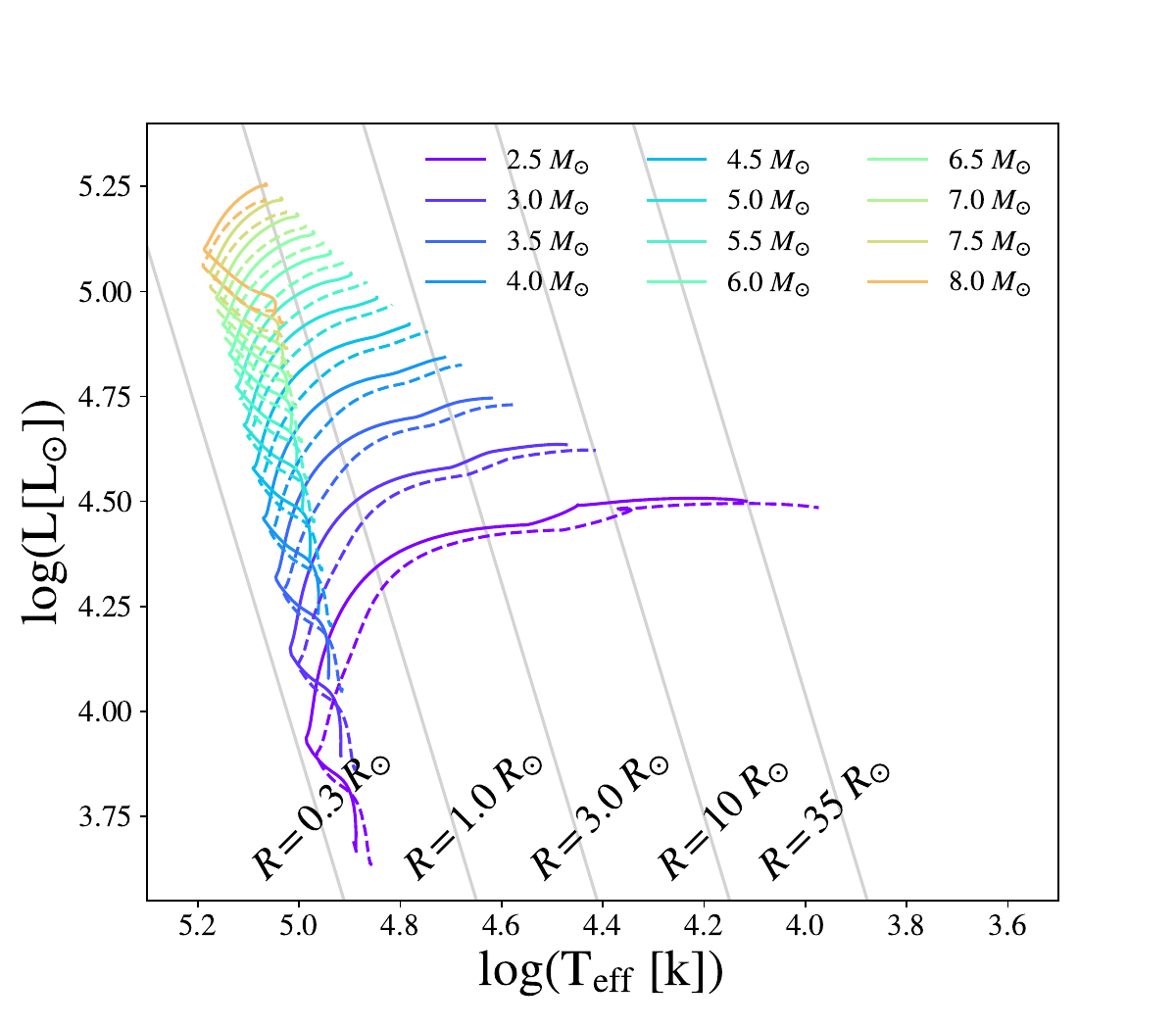}
     \caption{Hertzsprung-Russell diagram of single He-rich stars with initial mass from 2.5 $M_{\odot}$ to 8.0 $M_{\odot}$ evolving from ZAHeMS to their central carbon exhaustion (solid lines: non-rotating models, dashed lines: rotating models). The grey lines refer to contours of constant radii.}
     \label{fig1}
\end{figure}

\begin{figure}
     \centering
     \includegraphics[width=1.0\columnwidth]{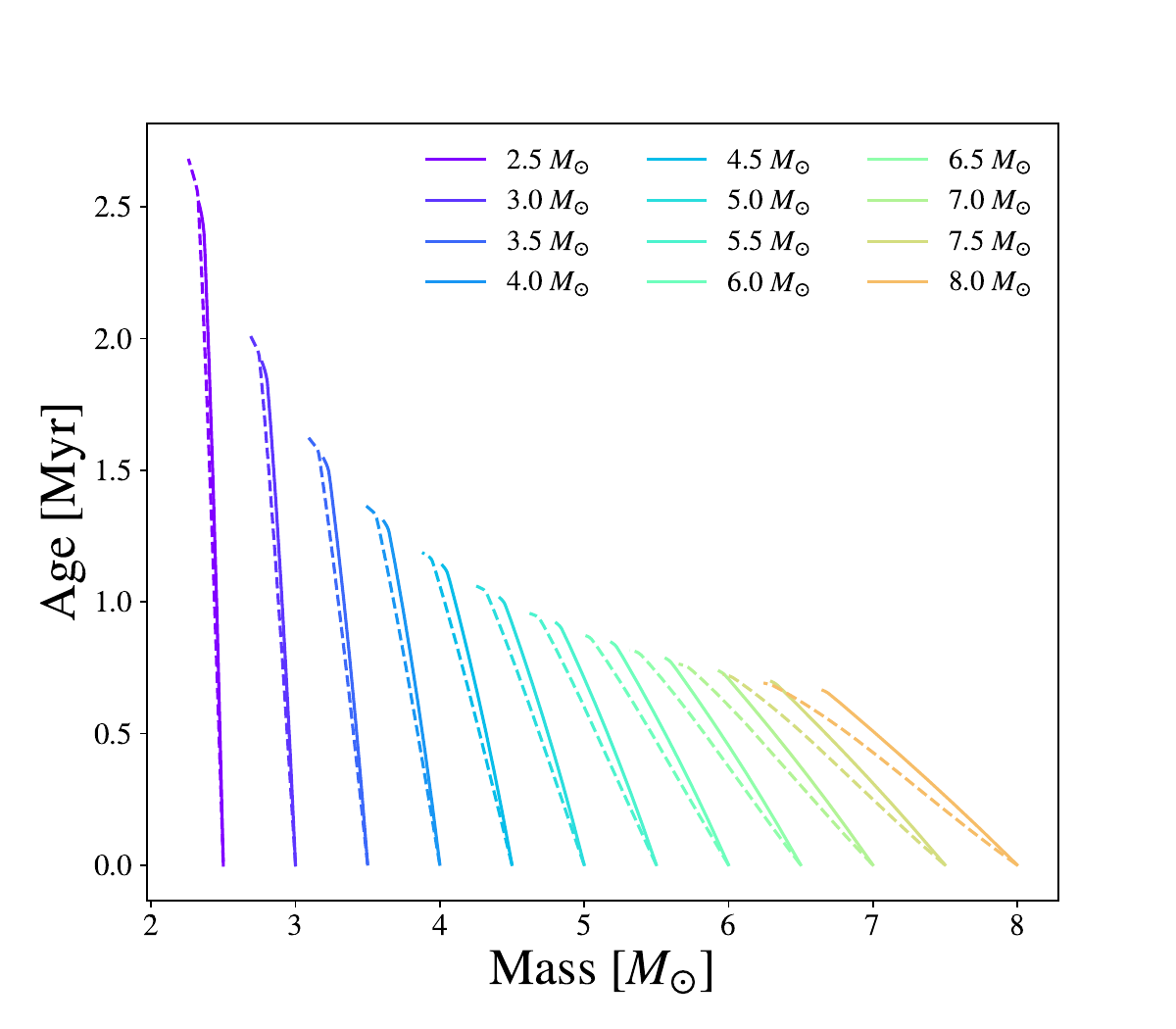}
     \caption{Same models as in Figure \ref{fig1}, we show ages from the onset of the core helium burning to their central carbon depletion as a function of their masses.}
     \label{fig2}
\end{figure}

\begin{figure}
     \centering
     \includegraphics[width=1.0\columnwidth]{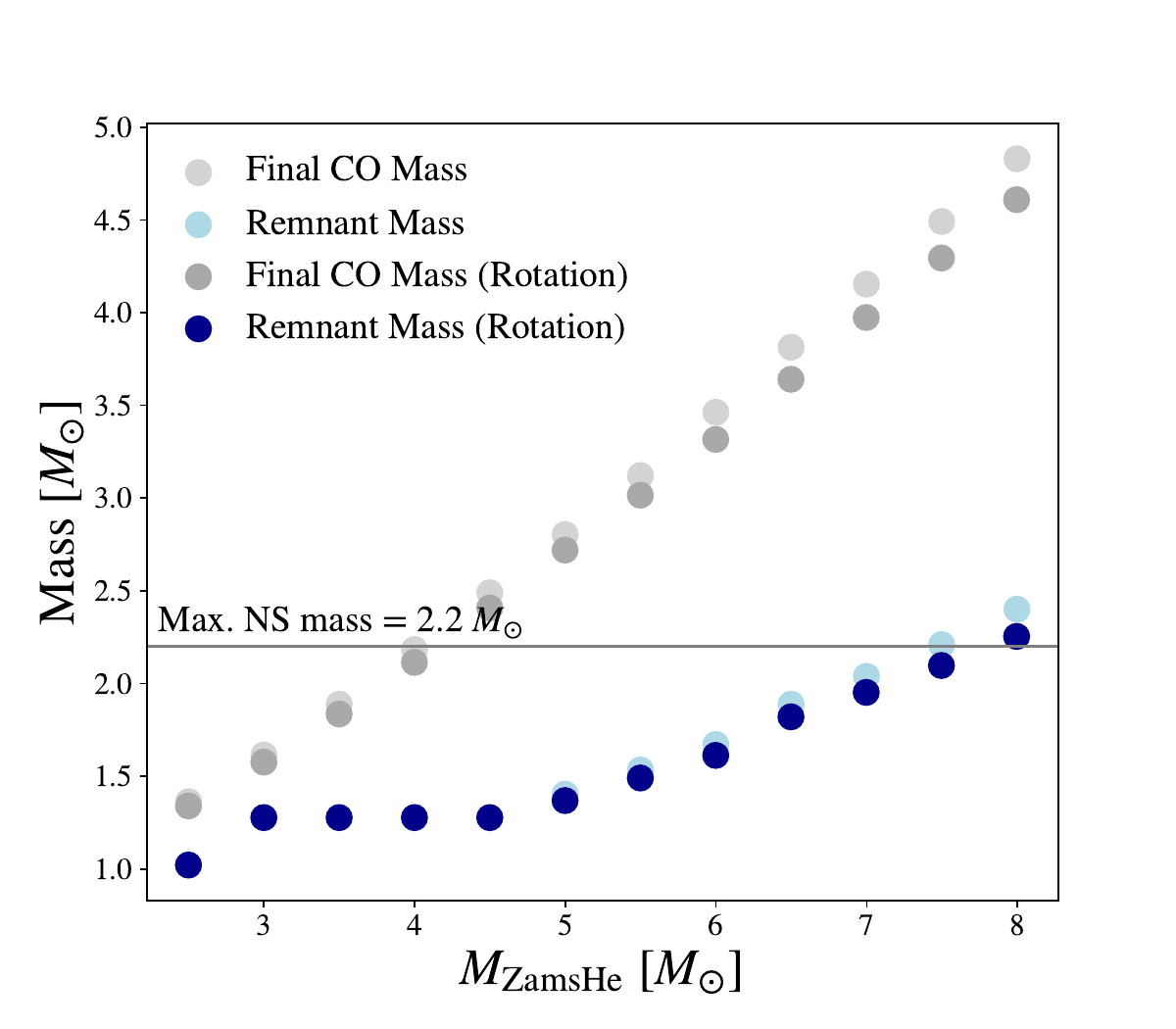}
     \caption{The carbon-oxygen (CO) core mass of He-rich stars at the end of central carbon depletion (Final CO core mass) and remnant mass as a function of their initial mass $M_{\rm ZamsHe}$. The symbols in dark color mark the models with rotation.}
     \label{fig3}
\end{figure}

In this section, we introduce the modeling of rotating and non-rotating He-rich stars. Figure \ref{fig1} presents the Hertzsprung-Russell (HR) diagram of He-rich stars from the onset of core helium burning (i.e., ZAHeMS) to the carbon depletion in the center. We select He-rich stars from 2.5 to 8.0 $M_{\odot}$ with their initial metallicity $Z = Z_{\odot}$. Solid lines represent the evolutionary tracks of non-rotating models, while the dashed ones refer to those of fast-rotating models with their initial surface rotation velocity of 600 $\rm  km$ $\rm s^{-1}$. During the core helium burning phase, stars first move to the upper left of the HR diagram, increasing both the effective temperature and the luminosity. Then subsequent evolution brings the stars to the right especially for low-mass stars, decreasing the temperature due to their expansion during the post-main-sequence phase. In contrast, rotating stars evolve like being less massive (see dashed lines in Figure \ref{fig1}), as the rotation can reduce the surface gravity. Rotation is also considered to have two additional effects. First, stars with rotation tend to lose more mass due to enhanced wind mass-loss rate \citep{Langer1998}. Additionally, rotation can trigger chemical mixing \cite[e.g.,][]{Heger2000}, which produces a larger core in the star. Therefore, rotating stars tend to have a longer lifetime (see Figure \ref{fig2}). In Figure \ref{fig3}, we present the baryonic remnant mass calculated following the ``\texttt{delayed}'' supernova prescription in \cite{Fryer2012}. We note that rotating models of He-rich stars with initial mass less than around 6 $M_{\odot}$ result in relatively smaller CO mass and thus remnant mass. This is because massive stars have strong winds driven by high radiation pressure and thus significant amounts of mass-loss. Furthermore, an NS is produced for He-rich stars with initial mass $M_{\rm ZamsHe}\lesssim 8\,M_\odot$, beyond which a BH is formed.

\section{Detailed evolution of He-rich stars in close orbits} \label{sec:4}

\subsection{Mass transfer under Eddington-limited and Super-Eddington accretion} \label{sec:4.1}
\begin{figure}
     \centering
     \includegraphics[width=\columnwidth]{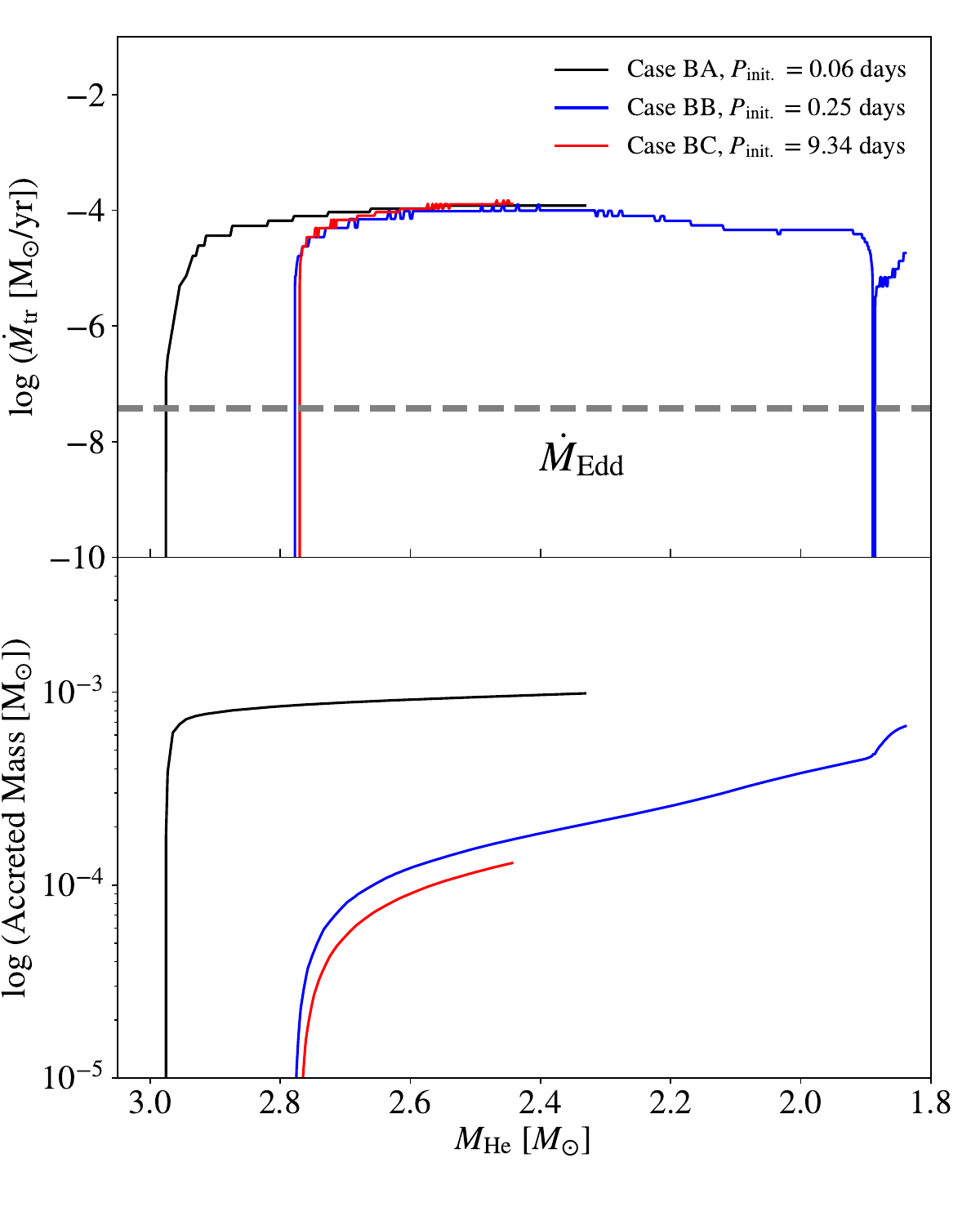}
     \caption{MT rate (upper panel) and accreted mass (lower panel) onto the first-born NS as a function of primary star mass. We select a He-rich star with its initial mass of $M_{\rm ZamsHe}$ = 3.0 $M_{\odot}$ at three different initial orbital periods (0.06 days, 0.25 days, and 9.34 days). The corresponding MT cases (Case BA: black; Case BB: blue; Case BC: red) are marked in different colors. The grey dashed line represents the standard Eddington accretion rate (1 $\dot{M}_{\rm Edd}$).}
     \label{fig4}
\end{figure}

He-rich stars in close binary systems are found to interact with their companions \citep[e.g., ][]{Ivanova2003,Tauris2015}. In Figure \ref{fig4}, we show various interactions of detailed binary evolution modeling of three binary sequences, consisting of a He-rich star with the initial mass of $M_{\rm ZamsHe}$ = 3.0 $M_{\odot}$ and an NS companion at different initial orbital periods (i.e., $P_{\rm init}$/days = 0.06, 0.25, and 9.34). Here we assume an Eddington-limited accretion ($\dot{M}_{\rm Edd}$) between a He-rich star and an NS. In the upper panel, at an initial orbital period of $P_{\rm init}$ = 0.06 days, MT initiates from the He-rich star during the core-helium burning phase (i.e., so-called Case BA MT phase) onto its companion (see black solid line). Under Eddington-limited accretion, the accreted mass onto the NS is $\sim$ 9.9 $\times$ 10$^{-4}$ $M_\odot$ (see black solid line in the bottom panel). For systems with initially wide orbits, the MT starts with late nuclear burning stages, i.e., Case BB (MT starting with shell-helium burning stage) for $P_{\rm init}$ = 0.25 days and Case BC (MT starting with core-carbon burning stage) for $P_{\rm init}$ = 9.34 days. The accreted mass onto the NS is $\sim$ 6.7 $\times 10^{-4}$ $M_\odot$ for Case BB and $\sim$ 1.3 $\times 10^{-4}$ $M_\odot$ for Case BC, respectively. 

\begin{figure}
     \centering
     \includegraphics[width=\columnwidth]{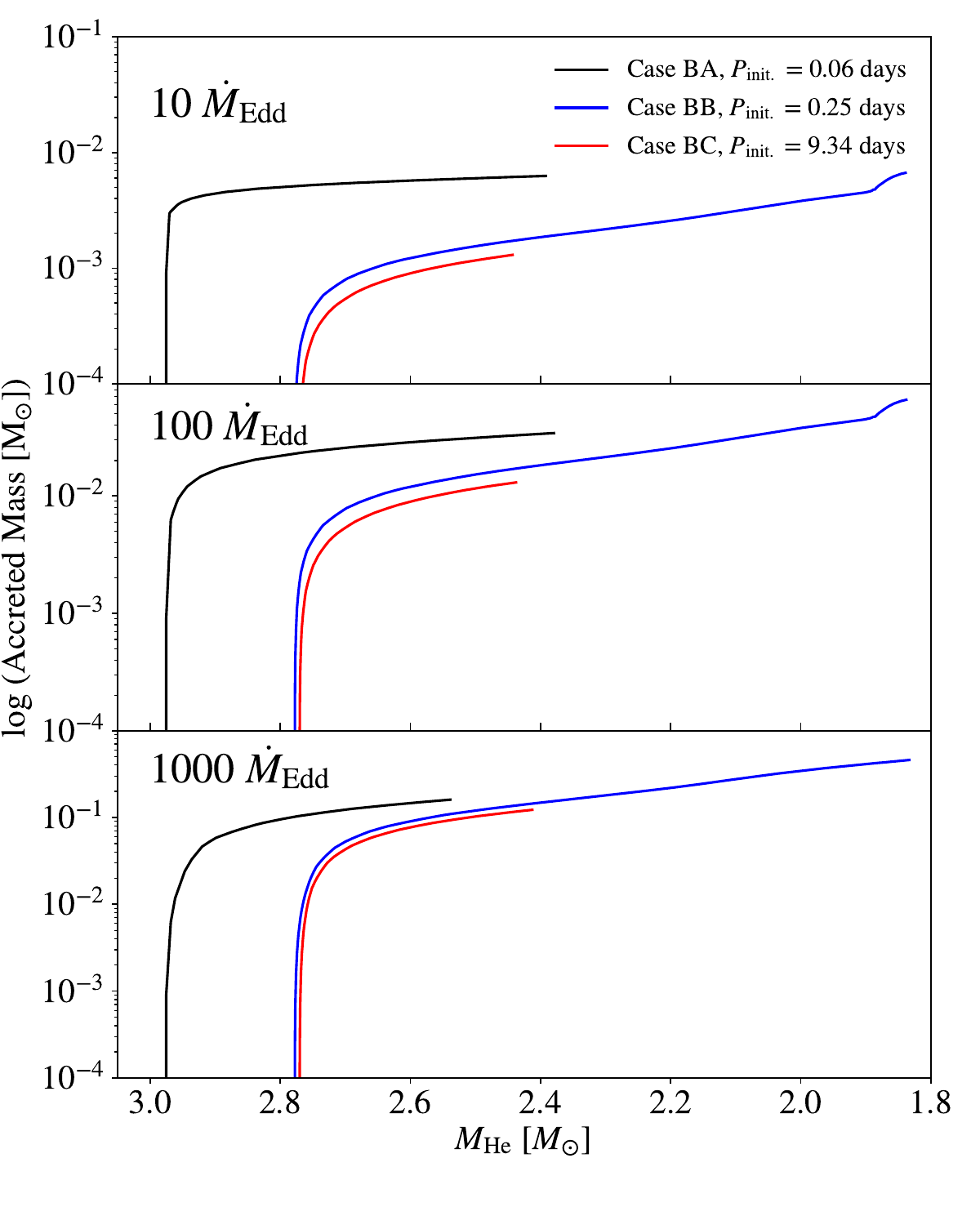}
     \caption{Similar to the lower panel of Figure \ref{fig4}, but with different Eddington accretion rate limits (Upper panel:$10\,\dot{M}_{\rm Edd}$; Middle panel: $100\,\dot{M}_{\rm Edd}$; Bottom panel: $1,000\,\dot{M}_{\rm Edd}$).}
     \label{fig5}
\end{figure}

In contrast, Figure \ref{fig5} shows the accreted mass onto the NS for the same binary sequences as above but with different Eddington accretion limits. In the case of 10 $\dot{M}_{\rm Edd}$ (see the upper panel), an NS can accrete mass of $\sim$ 6.3 $\times 10^{-3}$ $M_\odot$ for Case BA, $\sim$ 6.7 $\times 10^{-3}$ $M_\odot$ for Case BB, and $\sim$ 1.3 $\times 10^{-3}$ $M_\odot$ for Case BC, respectively. This finding shows that the accreted mass is around one order of magnitude higher when compared with the Eddington-limited accretion. When assuming a 100 Eddington accretion rate (see the middle panel), an NS can accrete more mass from its He-rich star companion for Case BA ($\sim$ 3.4 $\times 10^{-2}$ $M_\odot$), Case BB ($\sim$ 6.6 $\times 10^{-2}$ $M_\odot$), and Case BC ($\sim$ 1.3 $\times 10^{-2}$ $M_\odot$). Considering an even higher Eddington accretion rate (i.e., $1,000\,\dot{M}_{\rm Edd}$, see the bottom panel), one can see that the accreted mass onto the NS is apparently higher ( Case BA: $\sim 0.16$ $M_\odot$; Case BB: $\sim 0.46$ $M_\odot$ and Case BC: $\sim 0.12$ $M_\odot$). Note that the accreted mass from Case BA is, however, less, as the duration of the MT for these systems is relatively short due to reaching the limit of $\dot{M}_{\rm crit}$.

\begin{figure*}
     \centering
     \includegraphics[width=2\columnwidth]{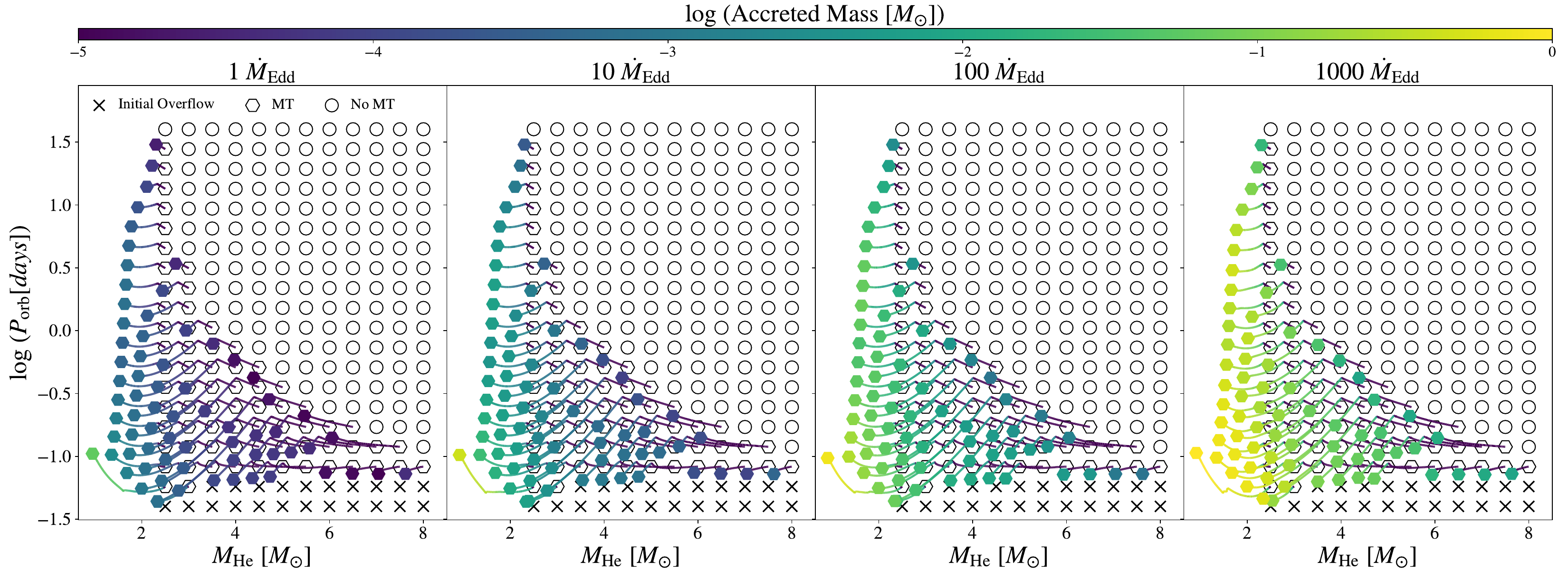}
     \caption{Accreted mass onto the first-born NS as a function of the He-rich star mass $M_{\rm He}$ and the orbital period $P_{\rm orb}$. Various symbols refer to different interactions: cross: initial overflow; circle: no MT; hexagon: MT.} 
     \label{fig6}
\end{figure*} 

\begin{figure*}
     \centering
     \includegraphics[width=2\columnwidth]{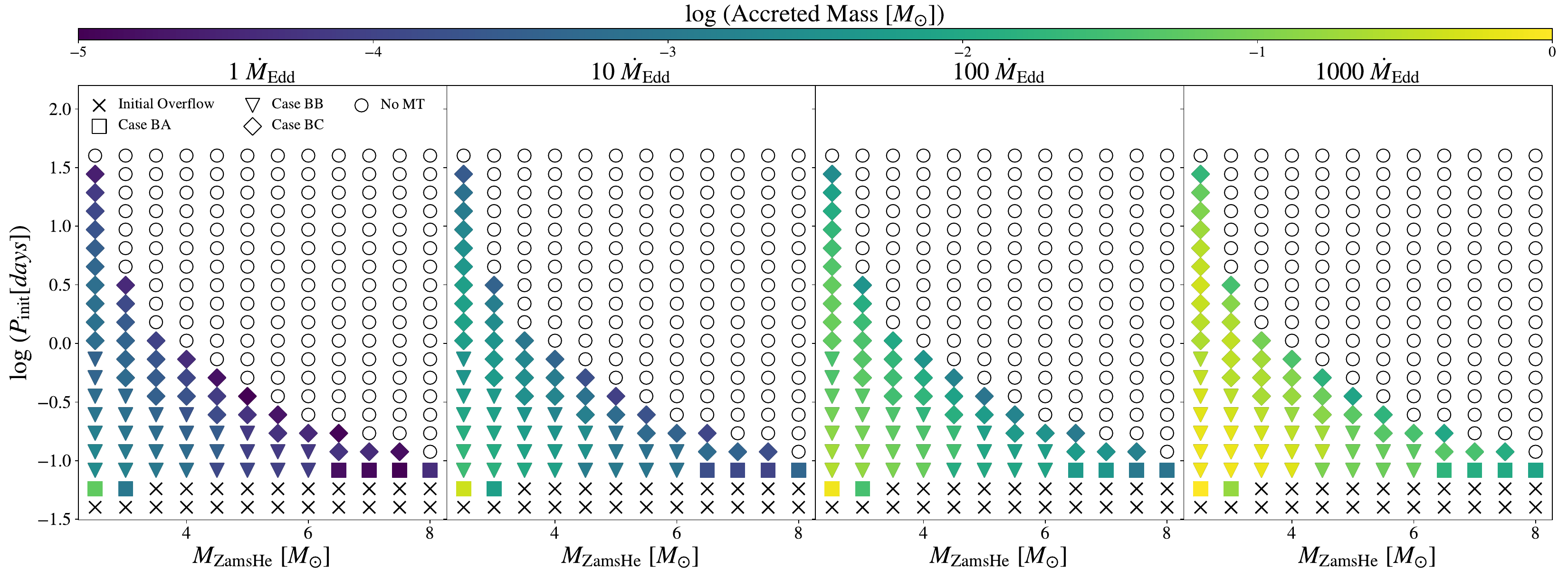}
     \caption{Final accreted mass onto the first-born NS as a function of an initial He-rich star mass $M_{\rm ZamsHe}$ and an initial orbital period $P_{\rm init}$. Various symbols indicate different interactions: cross: initial overflow; square: Case BA MT; triangle: Case BB MT; circle: Case BC MT.} 
     \label{fig7}
\end{figure*} 

\subsection{Parameter space analysis} \label{sec:4.2}
In this section, we investigate how a He-rich star interacts with its companion (1.33 $M_\odot$ NS) under different initial conditions, i.e., the initial mass of a He-rich star ($M_{\rm ZamsHe}$) and an initial orbital period ($P_{\rm init}$). Furthermore, we adopt four different Eddington accretion limits ($1\,\dot{M}_{\rm Edd}$, $10\,\dot{M}_{\rm Edd}$, $100\,\dot{M}_{\rm Edd}$, and $1,000\,\dot{M}_{\rm Edd}$). We choose $M_{\rm ZamsHe}$ in a range of 2.5 - 8.0 $M_\odot$ linearly with a step $\Delta M$ = 0.5 $M_\odot$, and $P_{\rm init}$ in a logarithmic space from 0.04 days to 40 days. It has been already found that the initial rotation of massive He-rich stars has an impact on their final mass and angular momentum \cite[see Section 5.1.2 in][]{Qin2018}. For simplicity, in our binary modeling, He-rich stars of $M_{\rm ZamsHe} \leqslant 8 M_\odot$ were initially assumed to be synchronized with their orbits.

As presented in Figure \ref{fig1}, low-mass He-rich stars initially (at ZAHeMS) have smaller radii and thus can be fitted in shorter orbits. Additionally, these stars expand significantly after leaving the helium main sequence. Figure \ref{fig6} shows that no mass exchange occurs for all He-rich stars considered in this investigation when the initial orbital period is longer than $\sim$ 40 days. Below this limit, mass interactions via the first Lagrangian point ($L_1$) are expected within a given range of parameter space. The colored lines in Figure \ref{fig6} depict the evolution of the mass accreted onto NSs as the systems evolve. We are only focused on the systems that have mass interactions (see the hexagon symbols). Let us start with the first panel from the left, which refers to the standard Eddington-limited accretion assumption. We note that stars first move to the left due to losing mass. Winds removing mass from He-rich stars widen the system, while mass transfer from the high-mass component to the low-mass one shrinks the orbit. Therefore, whether the stars move up or down mainly depends on the competition between the wind mass-loss and the mass transfer. Moving to the other panels with high Eddington limits, we can see that the parameter space of mass interactions is the same, but the accreted mass increases as higher Eddington limits are considered. At the standard Eddington limit (see the leftmost panel), at most NSs can accrete $\sim$ 0.06 $M_\odot$. In contrast, given 1, 000 times Eddington limit, the accreted mass can even reach around 1.0 $M_\odot$ (see the rightmost panel).

In Figure \ref{fig7}, we present the final accreted mass onto NSs depending on different MT phases based on the evolutionary stages of the He-rich stars. In the left panel, we note that Case BC (see diamond symbols) MT occurs for the systems with an initially higher $M_{\rm ZamsHe}$ and a longer $P_{\rm init}$ than Case BB (see triangle symbols). Case BA (see square symbols), however, is found to occur in a very limited initial parameter space (see cross symbols in Figure \ref{fig7}). For the initial He-rich stars of $M_{\rm ZamsHe}$ $\gtrsim$ 3.5 $M_\odot$ and their initial orbital period of $P_{\rm init}$ $\gtrsim$ 0.06 days, the He-rich stars initiate overflowing via $L_1$ during the core-helium burning phase (i.e., Case BA). It is worth noting that the systems undergoing Case BA can reach a very high MT rate ($\sim$ $\dot{M}_{\rm crit}$). We terminate these systems as they are likely to undergo a common envelope \citep{Ivanova2003}, which is consistent with previous findings \cite[see their Figure 18 in][]{Tauris2015}.

Under the Eddington-limited accretion, the left panel of Figure \ref{fig7} presents the amount of mass accreted onto an NS as a function of different initial conditions. We mark distinctive mass exchanges with different symbols (see the legends). We note that the accreted mass from $\sim$ 9.0 $\times$ $10^{-6}$ $M_\odot$ to $\sim$ 6.0 $\times$ $10^{-2}$ $M_\odot$, which corresponds to Case BC and Case BA MT, respectively. Assuming a moderate Eddington accretion limit (e.g., see 10 $\dot{M}_{\rm Edd}$ in the second panel of Figure \ref{fig7}), it is found that an NS can accrete higher mass than Eddington-limited accretion, from $\sim$ 8.7 $\times$ $10^{-5}$ $M_\odot$ for case BC up to $\sim$ 0.4 $M_\odot$ for Case BA. Under the assumption of 100 $\dot{M}_{\rm Edd}$ (see the third panel of Figure \ref{fig7}), we note that the maximum mass accreted onto an NS for Case BA can reach around 0.77 $M_\odot$ ($\sim$ 0.26 $M_\odot$ for Case BB and $\sim$ 0.07 $M_\odot$ for Case BC). In the last panel of Figure \ref{fig7} ($1,000\,\dot{M}_{\rm Edd}$), NSs with Case BA can accrete mass close to around 1.1 $M_\odot$, while the minimum mass accreted onto an NS is 8.3 $\times$ $10^{-3}$ $M_\odot$ for Case BC. 

\begin{figure*}
     \centering
     \includegraphics[width=2\columnwidth]{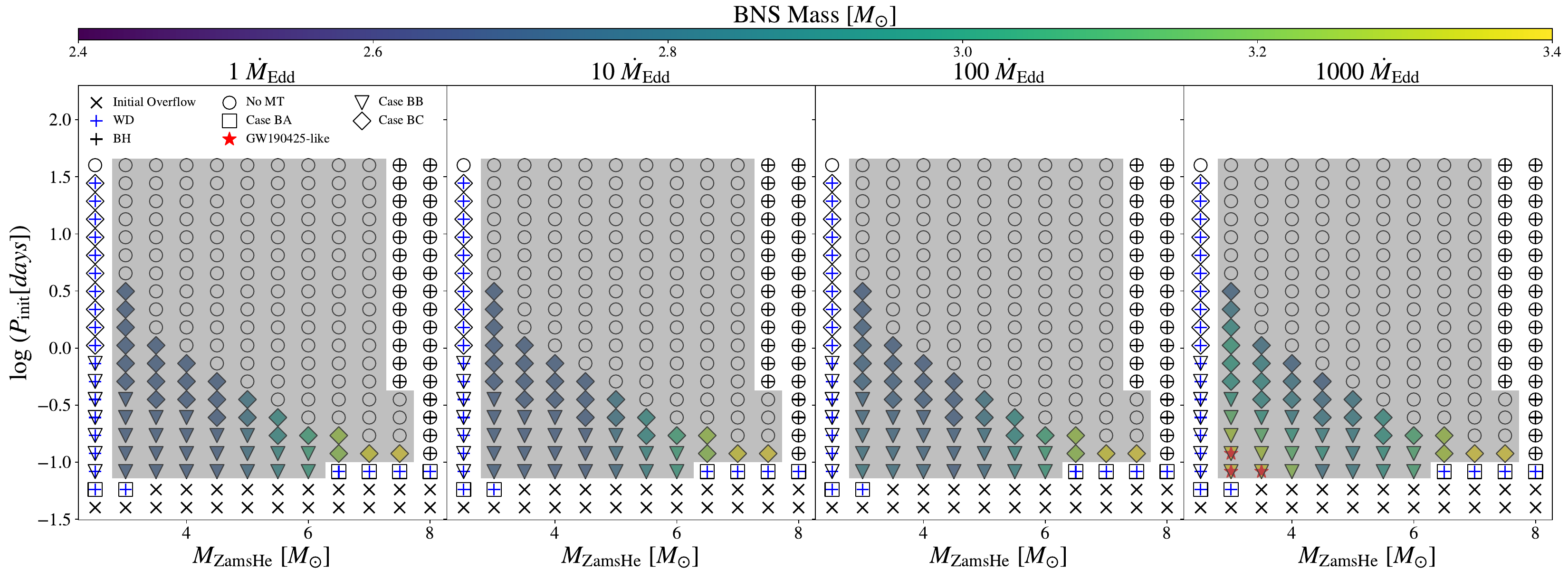}
     \caption{Similar to Figure \ref{fig7}, but the color bar refers to the total mass of the binary NSs. The blue and black plus symbol marks the He-rich stars that form white dwarfs and BHs, respectively. The red star symbols represent systems that could resemble GW190425-like events. The shaded area is marked for the parameter space in which a binary NS could be formed.} 
     \label{fig8}
\end{figure*} 

\begin{figure*}
     \centering
     \includegraphics[width=2\columnwidth]{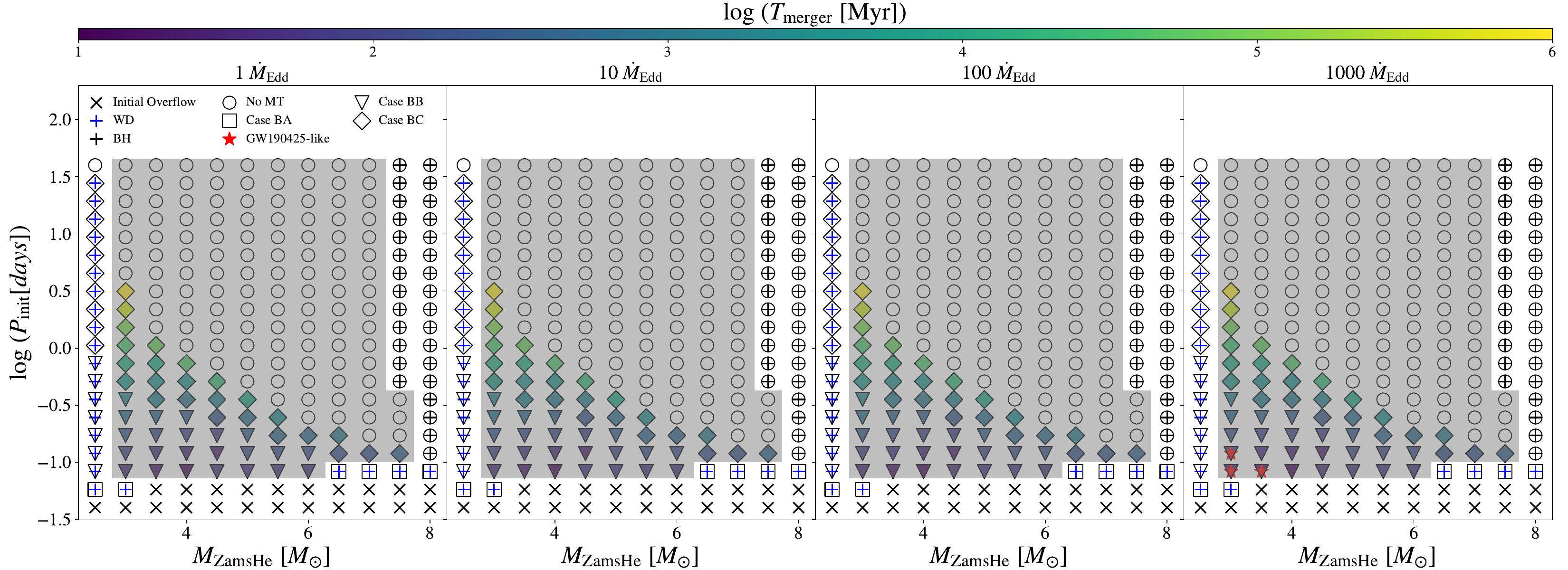}
     \caption{Similar to Figure \ref{fig7}, but the color bar refers to the merger times $T_{\rm merger}$.} 
     \label{fig9}
\end{figure*} 

After the depletion of the central carbon is reached, we follow the ``\texttt{delayed}'' supernova prescription \citep{Fryer2012} to calculate the NS mass. Figure \ref{fig8} presents the total mass of binary NSs in the MT of Case BA and Case BB. Here we assume that the first-born NS has a typical mass of 1.33 $M_\odot$ \citep{Lattimer2012}. We mark the parameter space with the shaded area where a BNS system can be formed. We only consider the BNS systems whose immediate progenitors have gone through MT. We note that no GW190425-like events could be produced when the accretion rate is not higher than 1, 000 times the Eddington limit. Intriguingly, GW190425-like events can be formed under the assumption of $1,000\,\dot{M}_{\rm Edd}$ (see the red stars in the rightmost panel). Besides, these systems are found to have gone through Case BB MT, with accreted mass from $\sim$ 0.70 $M_\odot$ to $\sim$ 0.77 $M_\odot$. In order to reproduce GW190425-like events, the immediate progenitors should have $M_{\rm ZamsHe}$ in a range of 3.0 - 3.5 $M_\odot$ and $P_{\rm init}$ from 0.08 days to 0.12 days. 

After a BNS system is formed, gravitational wave emission shrinks the separation by removing the orbital angular momentum and finally leads to the merger of the compact objects \citep{Peters1964}. As discussed in \cite{Tauris2015}, the supernova kicks of the second-born NS formed from ultra-stripped progenitors are expected to be small for two factors. First, it has been demonstrated \citep[see Section 4 in][]{Tauris2015} that the ejecta mass is extremely small when compared to standard SN explosions. This small amount of ejecta may produce a weak gravitational tug on the newborn NS and thus a small kick. Second, a weak outgoing shock can quickly lead to their ejection due to the low binding energies of the envelopes (e.g., a few $10^{49}$ erg), potentially preventing the built-up of the large anisotropies. Therefore, we calculate the merger times ($T_{\rm merger}$) of the BNS system in Figure \ref{fig9} without including the impact of the natal kick imparted onto the second-born NS. The merger times $T_{\rm merger}$ of these GW190425-like events are found (see the red star symbols in the rightmost panel in Figure \ref{fig9}) to be from $\sim$11 Myr to $\sim$190 Myr.

\subsection{Population properties}

\begin{figure}
     \includegraphics[width=1.0\columnwidth]{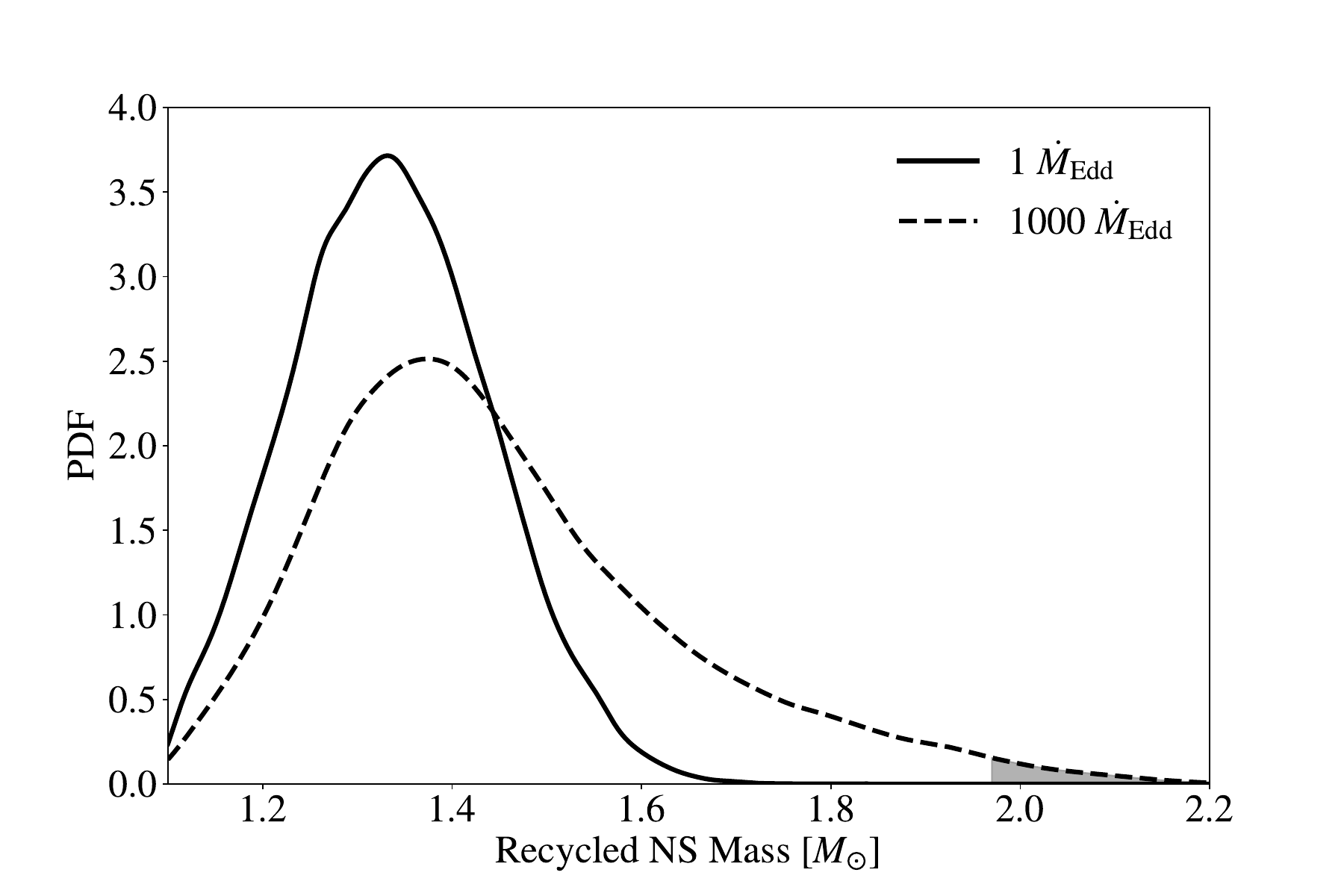}
     \caption{Probability density function (PDF) of the recycled (the first-born) NS mass under different Eddington accretion limits (Solid line: $1\,\dot{M}_{\rm Edd}$; dashed line: $1,000\,\dot{M}_{\rm Edd}$). The shaded area is marked for the mass range representing the recycled NS in GW190425-like systems.}
     \label{fig10}
\end{figure}

We further investigate the mass distribution of BNSs formed through the stable Case BB under the assumption of super-Eddington accretion. For simplicity, we adopt for the parameter space: i) initial mass distribution of helium stars is assumed to $dN \propto M^{-2.3}_{\rm ZamsHe}$ \cite[similar to the initial mass function as in][]{Kroupa2001}, $M_{\rm ZamsHe}/M_\odot \in [3.0, 7.5]$ uniformly in linear space; ii) $P_{\rm init} /{\rm days}$ $\in$ [0.05, 40] uniformly in logarithmic space; iii) first-born NS mass has a mass distribution consistent with that of Galactic BNSs, i.e., $M_{\rm NS}/M_{\odot} \sim \mathcal{N}(1.33, 0.11^2)$ \citep{Lattimer2012}. In our modeling, we generate 10$^{7}$ NS - He-rich star binaries and each system was assigned a redshift randomly drawn from the star formation rate described in \cite{Yuksel2008} following the simulation method from \cite{Sun2015,Zhu2021,Zhu2023}. Based on the results from our grids shown in Section \ref{sec:4.1} and \ref{sec:4.2}, we then use the multi-dimensional interpolation method to estimate the physical parameters, such as individual mass for two NSs, accreted mass $M_{\rm acc}$, merger time $T_{\rm merger}$. These parameters are assumed to be independent of redshift for simplicity. For each binary system, one can calculate the merger redshift since we have simulated its formation redshift and merger time. We select the BNS mergers that are formed within the local universe, i.e., redshift from $0 - 0.1$. In Figure \ref{fig10}, one can see the probability density function of the first-born NS mass under different accretion rates. In the case of $1,000\,\dot{M}_{\rm Edd}$ (1\,$\dot{M}_{\rm Edd}$), the first-born NSs have a Gaussian-like distribution with the median of $\sim$1.37\,$M_{\odot}$ (1.33\,$M_{\odot}$) and a long tail extending to $\sim$2.2\,$M_{\odot}$. Our simulation shows that GW190425-like events could contribute $\sim1.3\%$ of the total BNS mergers that are formed in the local Universe by adopting a Galactic-like NS mass distributioon for the first-born NSs.

\section{Conclusions and discussion} \label{sec:5}
Under the assumption that GW190425 is formed through the evolution of massive stars, isolated binaries in the field. The immediate progenitor of GW190425 is a close binary system composed of an NS and a He-rich star. In this paper, we explore the origin of GW190425 with detailed binary modeling, allowing for supper-Eddington accretion for NSs. We perform detailed stellar structure and binary evolution calculations that take into account wind mass-loss, internal differential rotation, and tidal interactions between the He-rich star and the NS companion. 

Starting with single, low-mass He-rich stars, we present how rotation can affect their evolution and final fates. We then systematically explore the parameter space of initial binary properties, including initial He-rich star masses and orbital periods. Additionally, assuming different Eddington accretion limits ($1\,\dot{M}_{\rm Edd}$, $10\,\dot{M}_{\rm Edd}$, $100\,\dot{M}_{\rm Edd}$, and $1,000\,\dot{M}_{\rm Edd}$) for the first-born NSs, we investigate their mass growth by accretion from the low-mass He-rich stars. We first identify the parameter space in which various MT phases can occur, consistent with earlier studies \citep{Ivanova2003,Tauris2015}. Given the parameter space investigated in this work, we further constrain the parameter space in which BNS systems can be formed. Under different Eddington accretion limits, we find that the accreted mass onto NSs in a range of $(\sim$ 9.0 $\times$ $10^{-6}$ $M_\odot$, $\sim$ 6.0 $\times$ $10^{-2}$ $M_\odot)$ for $1\,\dot{M}_{\rm Edd}$, $(\sim$ 8.7 $\times$ $10^{-5}$ $M_\odot$, $\sim$ 0.4 $M_\odot)$ for $10\,\dot{M}_{\rm Edd}$, $(\sim$ 8.3 $\times$ $10^{-4}$ $M_\odot$, $\sim$ 0.77 $M_\odot)$ for $100\,\dot{M}_{\rm Edd}$, and $(\sim$ 8.3 $\times$ $10^{-3}$ $M_\odot$, $\sim$ 1.1 $M_\odot)$ for $1,000\,\dot{M}_{\rm Edd}$, respectively. Assuming the first-born NS with a typical mass of 1.33 $M_\odot$ \citep{Lattimer2012}, we find that, in order to reproduce GW190425-like events, supper-Eddington accretion (e.g., $1,000\,\dot{M}_{\rm Edd}$) is required. Additionally, their potential immediate progenitors should have $M_{\rm ZamsHe}$ in a range of 3.0 - 3.5 $M_\odot$ and $P_{\rm init}$ from 0.08 days to 0.12 days. For the GW190425-like systems just after the formation, their merger times due to gravitational wave emission vary from $\sim$ 11 Myr to $\sim$ 190 Myr. By adopting a Galactic-like NS mass distribution for the first-born NSs and assuming an accretion rate of $1,000\,\dot{M}_{\rm Edd}$, our simplified population synthesis simulation shows that GW190425-like events could contribute $\sim$1.3\% to the total BNS mergers that are formed in the local Universe. In order to achieve the high event rate density of GW190425-like BNS mergers \citep[$250-2810\,{\rm Gpc}^{-3}\,{\rm yr}^{-1}$;][]{gw190425}, a stable Case BB MT rate of $1,000\,\dot{M}_{\rm Edd}$ for global BNS population in Universe is required.

In principle, an NS can increase its mass efficiently as long as super-Eddington accretion is allowed. Recent studies in \cite{Chashkina2017,Chashkina2019} show that the accretion rate can be highly enhanced when NSs have strong surface magnetic fields \citep[e.g.,][]{Gao2021}. Considering magnetic fields for NSs, \cite{Gao2022} systematically investigated the formation of mass-gap BH binaries formed via the accretion-induced collapse of NSs in low-mass X-ray binaries, finding that super-Eddington accretion plays a key role in the mass growth of NSs. We expect the current O4 Observing run will capture GW190425-like events or heavy BNSs with an extreme mass ratio, which could be likely formed through this channel.

\section*{Acknowledgements}
Y.Q. acknowledges the support from the Doctoral research start-up funding of Anhui Normal University and from the Key Laboratory for Relativistic Astrophysics at Guangxi University. This work was partially supported by the National Natural Science Foundation of China (grant No. 12065017, 12192220, 12192221, U2038106), the Natural Science Foundation of Universities in Anhui Province (grant No. KJ2021A0106), and the Natural Science Foundation of Anhui Province (grant No. 2308085MA29). J.P.Z. thank the COMPAS group at Monash University. Q.W.T. acknowledges support from the Natural Science Foundation of Jiangxi Province of China (grant No. 20224ACB211001). F.L. is supported by the Shanghai Post-doctoral Excellence Program and the National SKA Program of China (grant No. 2022SKA0130103). E.W.L. is supported by the National Natural Science Foundation of China (grant No. 12133003). 

\section*{Data Availability}
The data generated in this work will be shared upon reasonable request to the corresponding author.



\bibliographystyle{mnras}
\bibliography{main} 








\label{lastpage}
\end{document}